\documentclass[showkeys,showpacs,superscriptaddress,nofootinbib]{revtex4-2}
\usepackage{amsmath}
\usepackage{amssymb}
\usepackage{dutchcal}
\usepackage{graphicx}
\usepackage[normalem]{ulem}
\usepackage{xcolor}
\usepackage{float}
\usepackage{slashed}

\newcommand{\Tr}{{\rm Tr}}
\newcommand{\be}{\begin{equation}}
\newcommand{\ee}{\end{equation}}

\allowdisplaybreaks

\begin{document}

\title{Non-Abelian monopoles in modified gravity
}
\author{
Vladimir Dzhunushaliev
}
\email{v.dzhunushaliev@gmail.com}
\affiliation{
Department of Theoretical and Nuclear Physics,  Farabi University, Almaty 050040, Kazakhstan
}
\affiliation{
Institute for Experimental and Theoretical Physics, Farabi University, Almaty 050040, Kazakhstan
}
\affiliation{Academician J.~Jeenbaev Institute of Physics of the NAS of the Kyrgyz Republic, 265 a, Chui Street, Bishkek 720071, Kyrgyzstan}

\author{Vladimir Folomeev}
\email{vfolomeev@mail.ru}


\affiliation{Academician J.~Jeenbaev Institute of Physics of the NAS of the Kyrgyz Republic, 265 a, Chui Street, Bishkek 720071, Kyrgyzstan}

\begin{abstract}
Within modified gravity, we study static spherically and axially symmetric self-gravitating non-Abelian monopoles in $SU(2)$ Yang-Mills-Higgs theory.
By comparing these monopoles with those obtained in Einstein-Yang-Mills-Higgs theory, we identify the differences introduced 
by the modification of gravity and show that they can be quite significant for systems with strong Higgs self-coupling.
\end{abstract}

\pacs{}

\keywords{Non-Abelian monopole, modified gravity }
\date{\today}

\maketitle 

\section{Introduction}

Modified gravity theories (MGTs) constitute a natural generalization of general relativity (GR) and are widely used to describe 
gravitational phenomena beyond its standard domain of applicability. The particular interest in such theories 
is motivated both by fundamental considerations related to a possible ultraviolet completion of gravity and 
by phenomenological issues in cosmology and astrophysics. In this context, $f(R)$ gravity occupies an important place, 
in which the Einstein-Hilbert action is generalized to a functional of the scalar curvature 
$R$~\cite{Sotiriou:2008rp,DeFelice:2010aj,Nojiri:2010wj,Nojiri:2017ncd}. This theory represents 
one of the simplest  yet nontrivial extensions of GR.

$f(R)$ gravity models possess a number of properties that make them particularly attractive for cosmological applications. 
 In particular, they enable an inflationary stage in the early Universe without introducing an additional fundamental scalar field, 
 as occurs, for example, in the Starobinsky model $f(R)=R+\gamma R^2$~\cite{Starobinsky:1980te}. 
In addition, such theories can effectively describe the observed accelerated expansion of the present Universe, 
offering an alternative to the standard $\Lambda$CDM model and to the interpretation of the cosmological constant 
as dark energy~\cite{Sotiriou:2008rp,DeFelice:2010aj,Nojiri:2010wj,Nojiri:2017ncd}. Due to these properties, 
$f(R)$ gravity is actively employed in both analytical studies and numerical modeling of cosmological evolution. 
These theories are also used, for instance, in constructing models of compact, strongly gravitating objects such as neutron 
and quark stars~\cite{Astashenok:2014dja,Astashenok:2017dpo,Folomeev:2018ioy,Olmo:2019flu}.

On the other hand, non-Abelian gauge theories predict the existence of topologically nontrivial field configurations with finite energy. 
A classical example of such solutions is the ’t~Hooft--Polyakov magnetic monopole, which arises in $SU(2)$ Yang-Mills-Higgs theory 
under spontaneous symmetry breaking~\cite{tHooft:1974kcl,Polyakov:1974ek}. This object represents a regular solitonic configuration 
characterized by a nontrivial topology of the vacuum manifold and by the presence of a magnetic charge determined by the homotopy 
properties of the Higgs field at spatial infinity. The ’t~Hooft--Polyakov monopole plays an important role in quantum field theory and cosmology, 
particularly in scenarios of phase transitions in the early Universe and the formation of topological defects~\cite{Vilen94}.

The inclusion of gravitational interaction necessitates the consideration of a self-consistent set of Einstein-Yang-Mills-Higgs equations. 
Within the framework of GR, such systems have been extensively studied: static spherically and axially symmetric 
solutions have been constructed, their regularity, asymptotic behavior, and stability have been analysed, and the influence of gravity 
on the mass and internal structure of the monopole has been investigated 
(see, e.g., Refs.~\cite{Lee:1991vy,Breitenlohner:1991aa,Breitenlohner:1994di,Kleihaus:1997ic,Volkov:1998cc,Hartmann:2001ic} and references therein). 
In particular, it has been shown that for sufficiently strong gravitational coupling, qualitatively new effects may arise, 
including the formation of horizons and a transition to black hole configurations with non-Abelian fields~\cite{Lee:1991vy,Volkov:1989fi,Volkov:1990sva,Bizon:1990sr,Brihaye:1999nn}.

Nevertheless, despite significant progress in the study of both MGTs and topological solitons, their combined treatment 
remains underexplored. In the context of $f(R)$ gravity,  additional interest arises from the fact that this theory is dynamically 
equivalent to a scalar-tensor theory with an extra scalar degree of freedom~\cite{Sotiriou:2008rp,DeFelice:2010aj,Nojiri:2010wj}. 
This allows one to formulate the theory either in the Jordan frame, where the modification directly affects the gravitational sector, 
or in the Einstein frame, obtained via a conformal transformation of the metric. In the latter, the gravitational part takes the standard Einstein 
form, while the additional degrees of freedom appear as a scalar field with a nontrivial potential and a coupling to matter. 
The choice of frame can significantly affect the interpretation of physical quantities, such as the mass and energy of solitonic configurations, 
as well as the formulation of boundary conditions.

The analysis of monopole solutions within $f(R)$ gravity may be of interest for several reasons. First, the modification of the gravitational sector 
introduces additional degrees of freedom that can significantly affect the structure of solitonic solutions. Second, such studies allow for examining 
the robustness of monopole properties established in GR with respect to deformations of gravitational dynamics. 
Finally, in a cosmological context, it is important to understand how topological defects behave under conditions corresponding 
to alternative models of the early and present Universe.

In the present work, we consider the ’t~Hooft--Polyakov magnetic monopole with magnetic charges $n=1$ and $n=2$ within the Starobinsky model, 
focusing on the first of the directions outlined above. Since we compare the magnitudes of physical quantities, 
such as the monopole mass, in GR and in MGT, it is more convenient to work in the Jordan frame. 
To this end, the corresponding self-consistent sets of equations for the gravitational, gauge, and scalar fields are derived, 
and their static spherically and axially symmetric solutions are obtained numerically. The main focus is placed on a comparative 
analysis of the spatial distributions of the field functions and the masses of the configurations obtained in GR and in modified gravity.

\section{The model}

We consider the (3+1)-dimensional $SU(2)$ Yang-Mills-Higgs system within modified gravity
[we use natural units with $c=\hbar=1$ throughout and the metric signature is $(+,-,-,-)$]:
\begin{equation}
S=\int d^4x~\sqrt{-g}\left[
-\frac{f(R)}{16 \pi G}  - \frac{1}{2}\Tr (F_{\mu \nu} F^{\mu \nu})
    + \Tr (D_\mu \phi~ D^\mu \phi)
    - \frac{\lambda}{4} \Tr \left(
        \phi^2 - \phi_0^2
    \right)^2
    \, \right],
\label{lgr_10}
\end{equation}
where $G$ is the Newtonian gravitational constant,
$f(R)$ is an arbitrary nonlinear function of the scalar curvature $R$, $g$
denotes the determinant of the metric tensor,  and the field strength tensor
of the  gauge field $A_\mu=\frac12 A_\mu^a \tau^a $ is
$$
F_{\mu\nu}=\partial_\mu A_\nu - \partial_\nu A_\mu + ie [A_\mu, A_\nu] \, ,
$$
where $a=1, 2, 3$ is a color index, $\mu, \nu = 0, 1, 2, 3$ are spacetime indices, and
$\tau^a$ are the Pauli matrices.
The covariant derivative of the scalar field in adjoint representation
$\phi=\phi^a \tau^a$ is
$$
D_\mu \phi = \partial_\mu\phi + ie [A_\mu,\phi] ,
$$
where $e$ is the gauge coupling constant.
The scalar potential with a Higgs vacuum expectation value  $\phi_0$
breaks the $SU(2)$  symmetry down to $U(1)$ and the scalar self-interaction constant $\lambda$ defines the mass of the
Higgs field, $M_s=\sqrt \lambda \phi_0$. 

For our purposes, we represent the function $f(R)$  in the form $f(R)=R+\gamma h(R)$, where $h(R)$ is a new arbitrary function of $R$
and $\gamma$ is an arbitrary constant. When $\gamma=0$, one recovers Einstein's general relativity.
The corresponding gravitational field equations can be obtained by
varying the action \eqref{lgr_10} with respect to the metric, yielding:
\begin{equation}
\label{mod_Ein_eqs_gen}
\left(1+\gamma h_R\right) G_\mu^\nu-\frac{1}{2}\gamma\left(h-R\,h_R \right)\delta_\mu^\nu+
\gamma \left(\delta_\mu^\nu g^{\alpha \beta}-\delta_\mu^\alpha g^{\nu \beta}\right)\left(h_R\right)_{;\alpha;\beta}=8\pi G\left[ \left( T_{\mu}^\nu \right)_{\text{YM}} + \left( T_{\mu}^\nu \right)_{\phi}\right].
\end{equation}
Here $G_\mu^\nu\equiv R_\mu^\nu-\frac{1}{2}\delta_\mu^\nu R$ is the Einstein tensor, $h_R\equiv dh/dR$, and
the semicolon denotes the covariant derivative. In turn,  the pieces of the total energy-momentum tensor are
\be
\begin{split}
\left( T_{\mu}^\nu \right)_{\text{YM}} = & - F^a_{\mu \alpha} F^{a \nu}_{ \beta} g^{\alpha \beta}
    + \frac{1}{4}\delta_\mu^\nu  F^2\, ,\\
\left( T_{\mu}^\nu \right)_{\phi} = & D_\mu \phi^a D^\nu \phi^a
    - \delta_\mu^\nu \left[ \frac{1}{2}D_\alpha \phi^a D^\alpha \phi^a - \frac{\lambda}{4} \left(
        \phi^2  - \phi_0^2
    \right)^2\right]\, .
\label{EMT}
\end{split}
\ee
The corresponding matter field equations are obtained from the action \eqref{lgr_10} by varying with respect to $A_\mu$ and $\phi^a$, respectively,
\be
\begin{split}
D_\nu F^{a \nu \mu}& =  -e \epsilon^{abc} \phi^b D^\mu \phi^c     \, , \\
D_\mu D^\mu \phi^a &+ \lambda \phi^a \left(\phi^2  - \phi_0^2 \right)     =  0  .
\label{field_eqs}
\end{split}
\ee

\section{Equations and solutions}

Working within the model described above, in this section we present the general spherically and axially symmetric equations and solve them numerically
for various values of the system parameters.

\subsection{The Ansatz}

In this paper we work within the framework of the Starobinsky model~\cite{Starobinsky:1980te}, for which
\begin{equation}
\label{h_R_part_pow}
f=R+\gamma h(R)\equiv R+\gamma R^2.
\end{equation}

\subsubsection{Spherically symmetric case}
\label{ans_spher}

For the gauge and Higgs fields we employ the usual static spherically symmetric
hedgehog Ansatz \cite{tHooft:1974kcl,Polyakov:1974ek}
\be
A_0^a=0,\quad A_i^a= \varepsilon_{aik} \frac{r^k}{er^2} \left[ 1 - W(r) \right],\qquad
\phi^a=\frac{r^a}{e r}H(r)\,.
\label{fields-boson}
\ee 

For the line element we employ isotropic coordinates
\begin{equation}
    ds^2 = f(r) dt^2 - \frac{l(r)}{f(r)} \left[dr^2+ r^2 \left(d\theta^2 + \sin^2 \theta d\varphi^2\right) \right].
\label{metric_sph}
\end{equation}
The metric function $f(r)$ can be rewritten as $f(r)=1-\frac{2 G \mu(r)}{r}$ with the mass function $\mu(r)$;  the ADM mass of the
configuration is defined as $M=\mu(\infty)$.

\subsubsection{Axially symmetric case}

To construct static axially symmetric solutions,  we use  isotropic coordinates with the spacetime metric in the Lewis-Papapetrou form
\begin{equation}
	ds^2 = f dt^2-\frac{m}{f}\left(d r^2+r^2 d\theta^2\right)-\frac{l}{f}r^2\sin^2\theta d\varphi^2,
\label{metric_axi}
\end{equation}
where the metric functions $f, l$, and $m$ depend on $r$ and $\theta$ only. The $z$-axis ($\theta=0$) represents the symmetry axis of the system.
Asymptotically (as $r\to \infty$), the functions $f, m, l \to 1$; i.e., the spacetime approaches a flat, Minkowski spacetime.

We take a purely magnetic gauge field, $A_0=0$,
and choose for the gauge field the Ansatz \cite{Rebbi1980,Hartmann:2001ic}
\begin{equation}
A_\mu dx^\mu =
-\frac{1}{2 e r} \left\{ \tau^n_\varphi 
 \left[ H_1 dr + \left(1-H_2\right) r d\theta \right]
 -n \left[ \tau^n_r H_3 + \tau^n_\theta \left(1-H_4\right) \right]
  r \sin \theta d\varphi \right\} . 
\label{ans_axi} 
\end{equation}
Here the symbols $\tau^n_r$, $\tau^n_\theta$, and $\tau^n_\varphi$
denote the dot products of the Cartesian vector
of the Pauli matrices, $\vec \tau = ( \tau_x, \tau_y, \tau_z) $,
with the spatial unit vectors
\begin{eqnarray}
\vec e_r^{\, n}      &=& 
(\sin \theta \cos n \varphi, \sin \theta \sin n \varphi, \cos \theta)
\ , \nonumber \\
\vec e_\theta^{\, n} &=& 
(\cos \theta \cos n \varphi, \cos \theta \sin n \varphi,-\sin \theta)
\ ,  \\
\vec e_\varphi^{\, n}   &=& (-\sin n \varphi, \cos n \varphi,0) 
\ , \nonumber
\label{rtp} 
\end{eqnarray}
respectively.
Since the fields wind $n$ times while the
azimuthal angle $\varphi$ covers the full trigonometric circle once,
the integer $n$ is referred to as the winding number of the solutions.
For the Higgs field, the corresponding Ansatz is chosen as follows~\cite{Rebbi1980,Hartmann:2001ic}:
\begin{equation}
\phi=\frac{1}{e} \left(H \tau_r^{n}+F \tau_\theta^{n}\right)  .
 \label{higgs} 
\end{equation}
The four gauge field functions $H_i$ 
and the two Higgs field functions $H$ and $F$ depend only on 
the coordinates $r$ and $\theta$.
For $n=1$ and $H_1=H_3=F=0$, $H_2=H_4=W(r)$, and $H(r,\theta)=H(r)$,
the axially symmetric Ansatz \eqref{ans_axi}--\eqref{higgs}  reduces to the spherically symmetric Ansatz~\eqref{fields-boson}.
Accordingly, the parameter $n$ determines the value of the magnetic charge, which is proportional to the topological charge: 
for $n=1$, we deal with a monopole solution with unit charge (spherically symmetric case), while for $n=2$, we deal with a multimonopole solution with charge 2 (axially symmetric case).

Finally, we fix the gauge by choosing the gauge condition as \cite{Hartmann:2001ic}
\begin{equation}
 r\, \partial_r H_1 - \partial_\theta H_2 = 0 
\ . \label{gc1} \end{equation}

\subsection{Equations and boundary conditions for the spherically symmetric case}

Substitution of
the Ansatz \eqref{h_R_part_pow}--\eqref{metric_sph}
into the general set of equations \eqref{mod_Ein_eqs_gen} and \eqref{field_eqs} yields the following
set of four coupled ordinary differential equations for the functions $W,H,f$, and $l$
(here the prime denotes differentiation  with respect to the radial coordinate, $f ^\prime = \frac{d f}{dx}$, etc.):
\begin{align}
    &
    \left(1+2\gamma R\right) f^{\prime\prime}-\frac{3}{4}\frac{f^{\prime 2}}{f}-\frac{f}{4}\left(\frac{ l^{\prime}}{l}\right)^2
    +\frac{1}{2}\left(\frac{4}{x}+\frac{l^\prime}{l}\right)f^\prime-\frac{f}{x}\frac{l^\prime}{l} \nonumber \\
    &
    +\alpha^2\left[\frac{\beta^2}{2}\left(1-H^2\right)^2 l-\frac{3}{x^4}\frac{f^2}{l}-\frac{2}{x^2}f H^2 W^2+
    \frac{3}{x^4}\frac{f^2}{l}\left(2-W^2\right)W^2+f H^{\prime 2}-\frac{2}{x^2}\frac{f^2}{l}W^{\prime 2}
    \right]  \nonumber\\
    &
    +\gamma\left[\frac{1}{2}\left(\frac{8}{x}-3\frac{f^\prime}{f}\right)R f^\prime+\left(f^\prime-\frac{2}{x}f\right)R\frac{l^\prime}{l}-\frac{1}{2}l R^2
    -\frac{1}{2}f R \left(\frac{l^\prime}{l}\right)^2+\left(3 f^\prime-f \frac{l^\prime}{l}\right)R^\prime+2 f R^{\prime\prime}
    \right]=0
    \, ,
\label{field_eqs_curved_Einstein_1}\\
    &
    \left(1+2\gamma R\right) l^{\prime\prime}+\frac{1}{2}l\left(\frac{f^\prime}{f}\right)^2+\left(\frac{1}{x}-\frac{l^\prime}{l}\right) l^\prime
    +\alpha^2\left\{\frac{\beta^2}{f}\left(1-H^2\right)^2 l^2+\frac{2}{x^4} f\left[\left(2-W^2\right) W^2-1\right]+2 l H^{\prime 2}
    \right\}\nonumber \\
    &
    +\gamma \left[l R\left(\frac{f^\prime}{f}\right)^2-\frac{l^2 R^2}{f}+2\left(\frac{1}{x}-\frac{l^\prime}{l}\right)R l^\prime+
    \frac{2}{x}\left(2 +x \frac{f^\prime}{f} \right)l R^\prime+4 l R^{\prime\prime}
    \right]=0
            \, ,
\label{field_eqs_curved_Einstein_2}\\
    &
    W^{\prime\prime}+\frac{1}{2} \left(2\frac{f^\prime}{f} - \frac{l^\prime}{l}\right)W^\prime -\frac{W^3}{x^2}+\left(\frac{1}{x^2}-\frac{l}{f}H^2\right)W    = 0 \, ,
\label{YM_eqs}\\
    &
    H^{\prime\prime} + \frac{1}{2}\left( \frac{4}{x} + \frac{l^\prime}{l} \right)H^\prime  -\frac{2}{x^2}H W^2 +\beta^2\frac{l}{f}\left(1-H^2\right)H  = 0 \, .
\label{phi_eqn}
\end{align}
In turn, the trace of Eq.~\eqref{mod_Ein_eqs_gen} gives the equation for the scalar curvature,
\begin{equation}
R^{\prime\prime}+\frac{1}{2}\left( \frac{4}{x} + \frac{l^\prime}{l} \right)R^\prime 
+\frac{1}{3\gamma}\left\{\frac{1}{2}\frac{l}{f}R
+\alpha^2\left[\beta^2\frac{l}{f}\left(1-H^2\right)^2+2\left(\frac{H W}{x}\right)^2+H^{\prime 2}
\right]
\right\}=0.
\label{scal_cur_eq}
\end{equation}
Equations \eqref{field_eqs_curved_Einstein_1}--\eqref{scal_cur_eq} are written in terms of the following dimensionless (tilded) variables and parameters:
a  radial coordinate, $x=e \phi_0 r$,  three rescaled  effective
coupling constants $\alpha^2=4\pi G\phi_0^2\,, \beta^2=\frac{\lambda}{e^2}\, , \tilde \gamma=(e \phi_0)^2 \gamma$,
the scalar field $\tilde H=H/(e \phi_0)$, and the scalar curvature $\tilde{R}=R/(e \phi_0)^2$
(to simplify the notation, we omit the tilde from now on).

The set of second-order ODEs \eqref{field_eqs_curved_Einstein_1}--\eqref{scal_cur_eq} is solved numerically subject to  boundary conditions derived from
the asymptotic expansion of the solutions at the boundaries of the integration domain, 
assuming regularity and asymptotic flatness.
 Explicitly, we impose:
\be
\begin{split}
\partial_x f(0)=&0, \quad \partial_x l(0)=0, \quad W(0)=1,\quad H(0)=0,\quad  \partial_x R(0)=0; \nonumber\\
f(\infty)=&1, \quad l(\infty)=1, \quad W(\infty)=0,\quad H(\infty)=1,\quad R(\infty)=0 \, . \nonumber
\end{split}
\ee

\subsection{Boundary conditions for the axially symmetric case}

Substituting the Ansatz \eqref{h_R_part_pow} and \eqref{metric_axi}--\eqref{higgs} into the general set of equations \eqref{mod_Ein_eqs_gen} and \eqref{field_eqs}
and taking into account the gauge condition (\ref{gc1}), one obtains a set of ten coupled elliptic partial differential equations 
for the functions $H, F, H_1,H_2,H_3,H_4, f, l, m$, and $R$. Using these equations, we  seek globally regular, asymptotically flat solutions with finite mass.
To avoid cluttering the text, we do not present these cumbersome PDEs here, providing only 
the appropriate boundary conditions for the gauge and scalar fields, metric functions, and scalar curvature imposed at the origin ($x=0$), at infinity ($x\to \infty$),
on the positive $z$-axis ($\theta=0$), and~--  exploiting the reflection symmetry with respect to $\theta\to \pi-\theta$~--
in the equatorial plane ($\theta=\pi/2$). Specifically, we require:
\begin{align}
&\left. \frac{\partial f}{\partial x}\right|_{x = 0} =
\left. \frac{\partial m}{\partial x}\right|_{x = 0} =
\left. \frac{\partial l}{\partial x}\right|_{x = 0} =
	\left. \frac{\partial R}{\partial x}\right|_{x = 0} =  0,  
\left. H_1\right|_{x = 0}= \left. H_3 \right|_{x = 0} = 0,  
\left. H_2\right|_{x = 0}= \left. H_4 \right|_{x = 0} = 1,  
\left. H\right|_{x = 0}=  \left. F \right|_{x = 0} = 0;\nonumber\\
&\left. f \right|_{x = \infty} = \left. m \right|_{x = \infty} =\left. l \right|_{x = \infty} =1 ,
	\left. R \right|_{x = \infty} = 0 ,
	\left. H_1\right|_{x = \infty}=\left. H_2 \right|_{x = \infty}=\left. H_3 \right|_{x = \infty}=\left. H_4 \right|_{x = \infty} =  0,
	\left. H\right|_{x = \infty} =  1,
	\left. F\right|_{x = \infty} =  0 ; \nonumber\\
&\left. \frac{\partial f}{\partial \theta}\right|_{\theta = 0,\pi/2} =\left. \frac{\partial m}{\partial \theta}\right|_{\theta = 0,\pi/2} =
\left. \frac{\partial l}{\partial \theta}\right|_{\theta = 0,\pi/2} =
	\left. \frac{\partial R}{\partial \theta}\right|_{\theta = 0,\pi/2} =  0 ,  \left. H_1\right|_{\theta = 0,\pi/2}=\left.H_3 \right|_{\theta = 0,\pi/2} = 0 ,\nonumber \\
&\left. \frac{\partial H_2}{\partial \theta}\right|_{\theta = 0,\pi/2} =\left. \frac{\partial H_4}{\partial \theta}\right|_{\theta = 0,\pi/2}=0,
\left. \frac{\partial H}{\partial \theta}\right|_{\theta = 0,\pi/2} =0, \left.F \right|_{\theta = 0,\pi/2} = 0. \nonumber
\end{align}
In turn,  the absence of a conical singularity requires the solutions to satisfy the constraint
 $\left. m\right|_{\theta=0}=\left. l\right|_{\theta=0}$ (this condition was verified during the calculations).

\subsection{Numerical method}

In order to map the infinite range of the radial variable $x$ 
to a finite interval, we introduce the compactified coordinate~$\bar x$ as follows:
\begin{equation}
     x = c_k\frac{\bar x}{1-\bar x^2} \, ,
\label{comp_coord}
\end{equation}
which maps the infinite region $[0;\infty)$ onto the finite interval $[0; 1]$. Here, $c_k$ is a constant used to adjust grid contraction.
In our calculations, we typically set $c_k=1$, though in some cases 
smaller values (down to 0.3) are chosen to obtain solutions for $\alpha$ approaching the critical value  $\alpha_{\text{cr}}$ (see below).

The equations are discretized on a grid, and the resulting set of nonlinear algebraic equations is solved using a modified Newton method. 
The underlying linear system is solved with the Intel MKL PARDISO sparse direct solver~\cite{pardiso} 
and the CESDSOL library\footnote{Complex Equations-Simple Domain 
partial differential equations SOLver, a C++ package developed by I.~Perapechka;
see Refs.~\cite{Herdeiro:2019mbz,Herdeiro:2021jgc}.}. For axially symmetric systems,
typical mesh sizes include $150\times 30$  points (up to $300\times 30$  points in some cases),
covering  the integration region $0\leq \bar x \leq 1$ 
[defined by the compactified radial coordinate~\eqref{comp_coord}] and $0\leq \theta \leq \pi/2$. 
For spherically symmetric systems, Eqs.~\eqref{field_eqs_curved_Einstein_1}--\eqref{scal_cur_eq}
are discretized on a grid of approximately 1000 grid points, with up to 3000 or more points used in certain cases. 
In all cases, the typical errors are on the order of $10^{-4}$.
The package provides an iterative procedure to obtain an exact solution starting from some initial guess configuration. 
For the latter, we use the gravitating monopole configuration found in Ref.~\cite{Breitenlohner:1991aa}.

\subsection{Results of numerical calculations}

In this subsection, we consider the dependence of gravitating spherically and axially symmetric (multi)monopole solutions with magnetic charge $n$ on the parameters
$\alpha, \beta$, and $\gamma$. In GR, when $\alpha$ increases from zero while $\beta$ is kept fixed,  fundamental branches of gravitating (multi)monopole solutions 
with charge $n$ emerge smoothly from the corresponding flat-space (multi)monopole solutions~\cite{Breitenlohner:1991aa,Hartmann:2001ic}. 
Depending on the Higgs self-coupling $\beta$,  there  exists a maximum value  $\alpha_{\text{max}}(n,\beta)$  beyond which solutions no longer exist,
both for the spherically  ($n=1$) and axially ($n=2$) symmetric cases.
However, in the Bogomolnyi-Prasad-Sommerfield (BPS) limit  ($\beta=0$) and for small $\beta$, the 
 $n=1$ case exhibits a short backward branch of unstable solutions. This branch extends down to a critical value
 $\alpha_{\text{cr}}(n,\beta)<\alpha_{\text{max}}(n,\beta)$, where the monopole branch reaches 
 a limiting solution and bifurcates with the branch of extremal Reissner-Nordstr\"{o}m (RN) black hole solutions~\cite{Breitenlohner:1991aa}.
For  $n=2$, such a backward branch is  absent, and $\alpha_{\text{max}}(n,\beta)$  coincides with $\alpha_{\text{cr}}(n,\beta)$~\cite{Hartmann:2001ic}.

Since this paper focuses on the influence coming from the modification of Einstein's gravity, 
here we study the corresponding changes in the total mass of the configurations as the modification of gravity is included.
The dimensionless ADM mass $\tilde M$ for both spherically and axially symmetric systems can be extracted from the metric function $f$ [see after Eq.~\eqref{metric_sph}] as follows:
\begin{equation}
\label{expres_mass}
\tilde M \equiv \frac{e}{4\pi \phi_0}M=\frac{1}{2\alpha^2}\lim_{x\to\infty} x^2\partial_x f = \frac{c_k}{4\alpha^2}\lim_{\bar x\to 1}\partial_{\bar x} f ,
\end{equation}
where  the last expression in the above equation represents the mass in terms of the
compactified coordinate $\bar x$ from Eq.~\eqref{comp_coord}.

Alternatively, the mass of the system in GR can be found using the Komar integral in the form
$$
\tilde M_K \equiv \frac{e}{4\pi \phi_0}M_K = \int \left(2\tilde{T}_0^0-\tilde{T}\right)\sqrt{-g}dx d\theta ,
$$
where $\tilde{T}_0^0$ and $\tilde{T}$ are the dimensionless energy density and the trace of the energy-momentum tensor, respectively.
This expression can be used to verify the numerical accuracy of the mass calculated via Eq.~\eqref{expres_mass}.
In addition, it is particularly useful for calculating the mass in the limit $\alpha \to 0$, where $\partial_{\bar x} f$ also vanishes.

\begin{figure}[t!]
\begin{minipage}[t]{.49\linewidth}
        \begin{center}
\includegraphics[width=1.\linewidth]{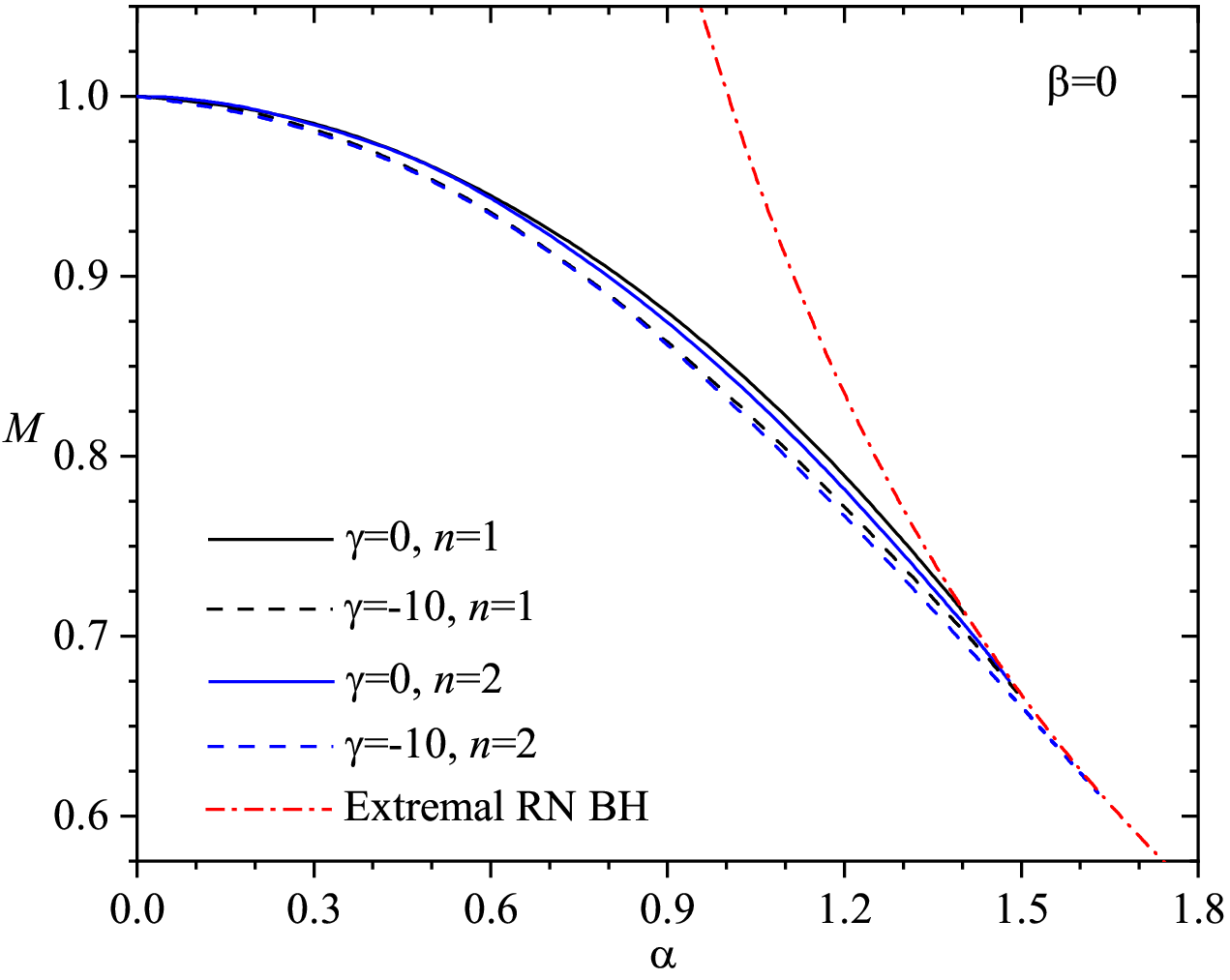}
        \end{center}
\end{minipage}\hfill
\begin{minipage}[t]{.49\linewidth}
        \begin{center}
\includegraphics[width=1.\linewidth]{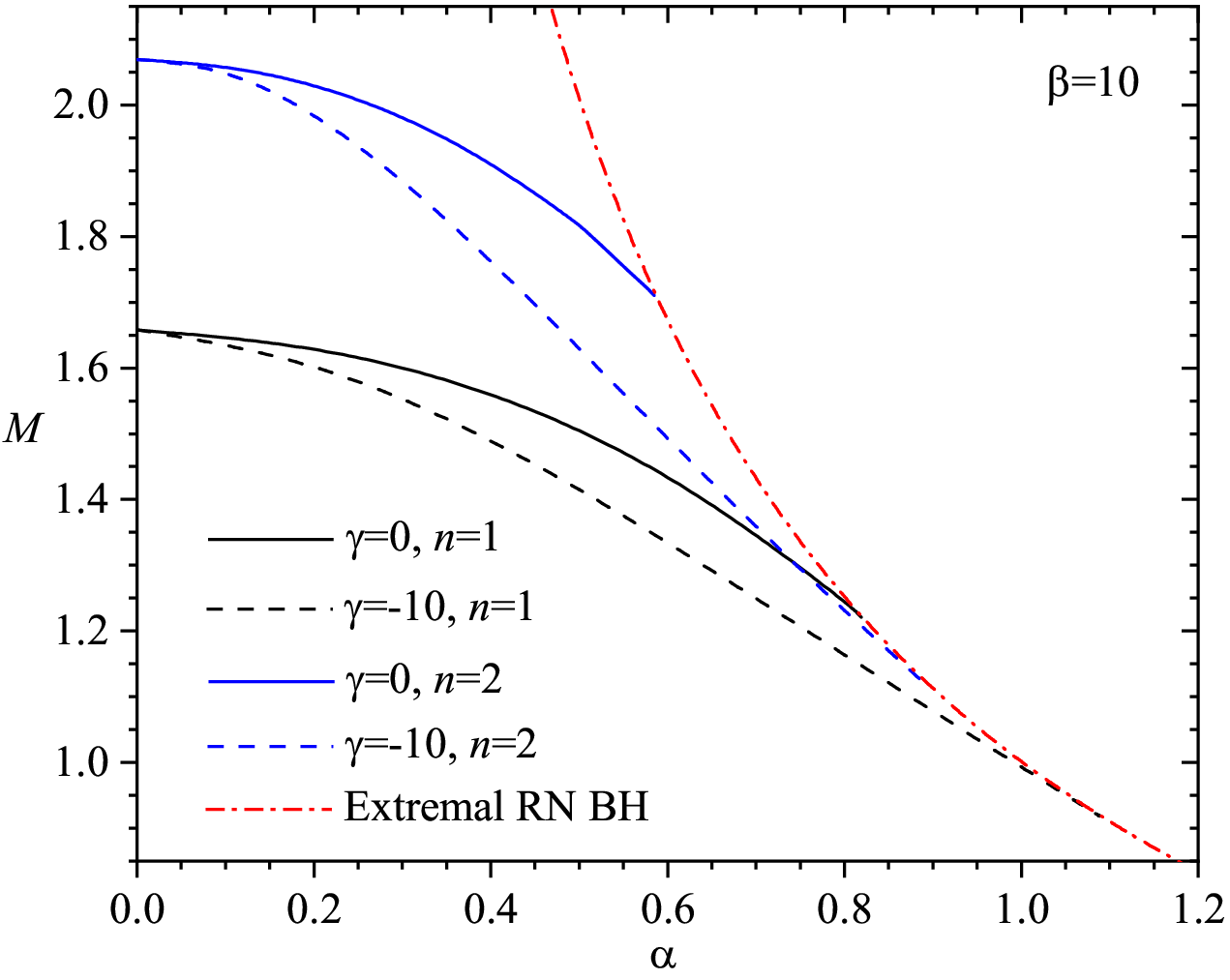}
        \end{center}
\end{minipage}\hfill
\caption{The dependence of the ADM mass $\tilde M$ from Eq.~\eqref{expres_mass} on the effective
gravitational coupling constant $\alpha$ is shown for $\beta=0, 10$ and $\gamma=0, -10$.
The solid lines correspond to the configurations in GR ($\gamma=0$), while the dashed lines represent the systems in MGT with $\gamma=-10$.
The black lines correspond to the monopole solutions ($n=1$), and the blue lines refer to the multimonopole solutions ($n=2$), for which the mass is given {\it per unit charge}.
For comparison, the mass of an extremal Reissner-Nordstr\"{o}m black hole of unit charge in GR~\cite{Hartmann:2001ic} is also shown.
}
\label{fig_mass_alph_n_1_2}
\end{figure}

\begin{figure}[t!]
\begin{center}
\includegraphics[width=1.\linewidth]{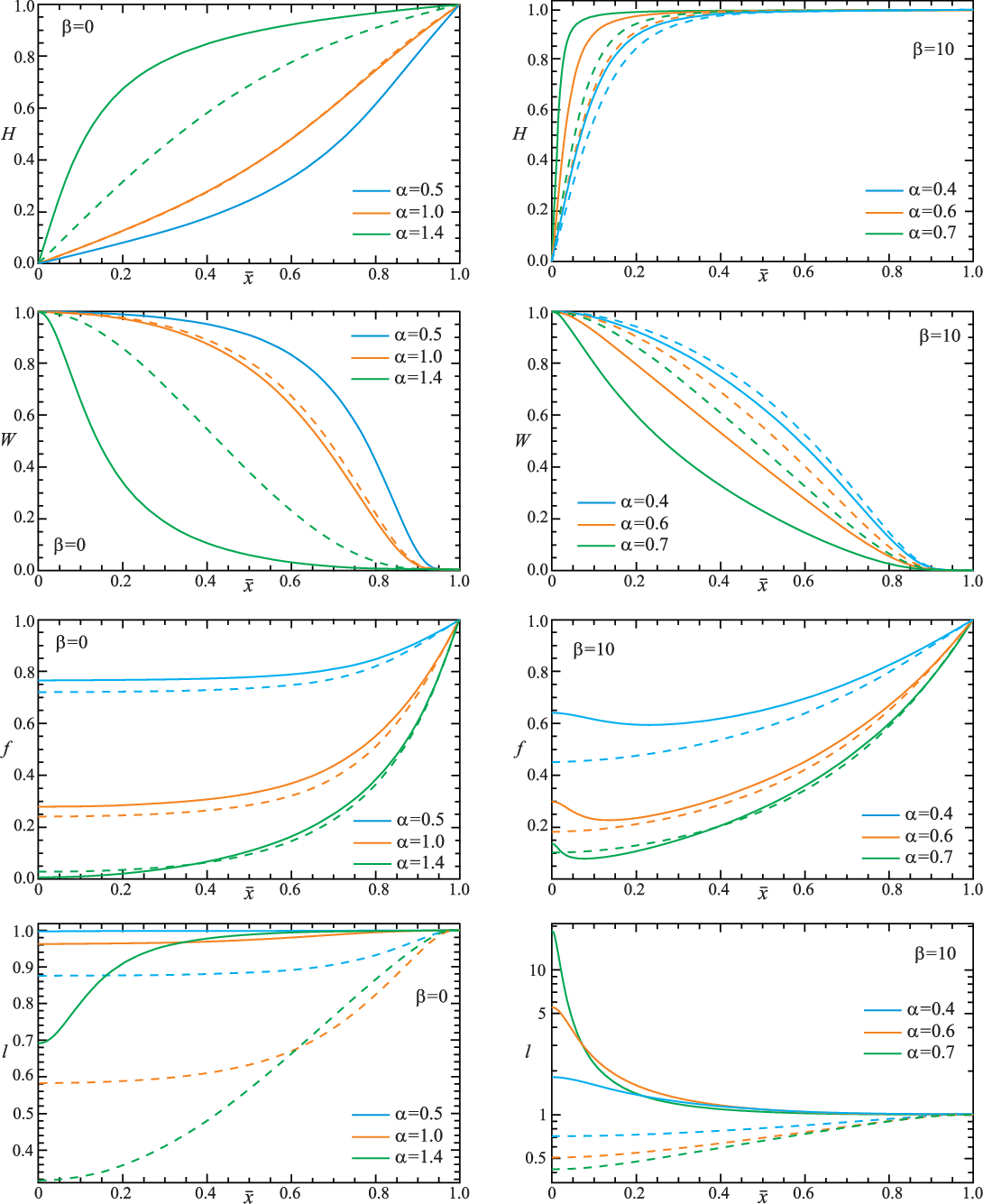}
\end{center}
\vspace{-0.5cm}
\caption{The spherically symmetric $n=1$ monopole solutions are shown for $\beta=0$ (left column) and $\beta=10$ (right column). The solid lines correspond to GR ($\gamma=0$),
while the dashed lines represent MGT with $\gamma=-10$. The plots are made using the compactified coordinate~$\bar x$ from \eqref{comp_coord} with $c_k=1$
(as is the case for Fig.~\ref{fig_sols_n_2_beta_0_10}). 
}
\label{fig_sols_n_1}
\end{figure}

\begin{figure}[t!]
\begin{center}
\includegraphics[width=1.\linewidth]{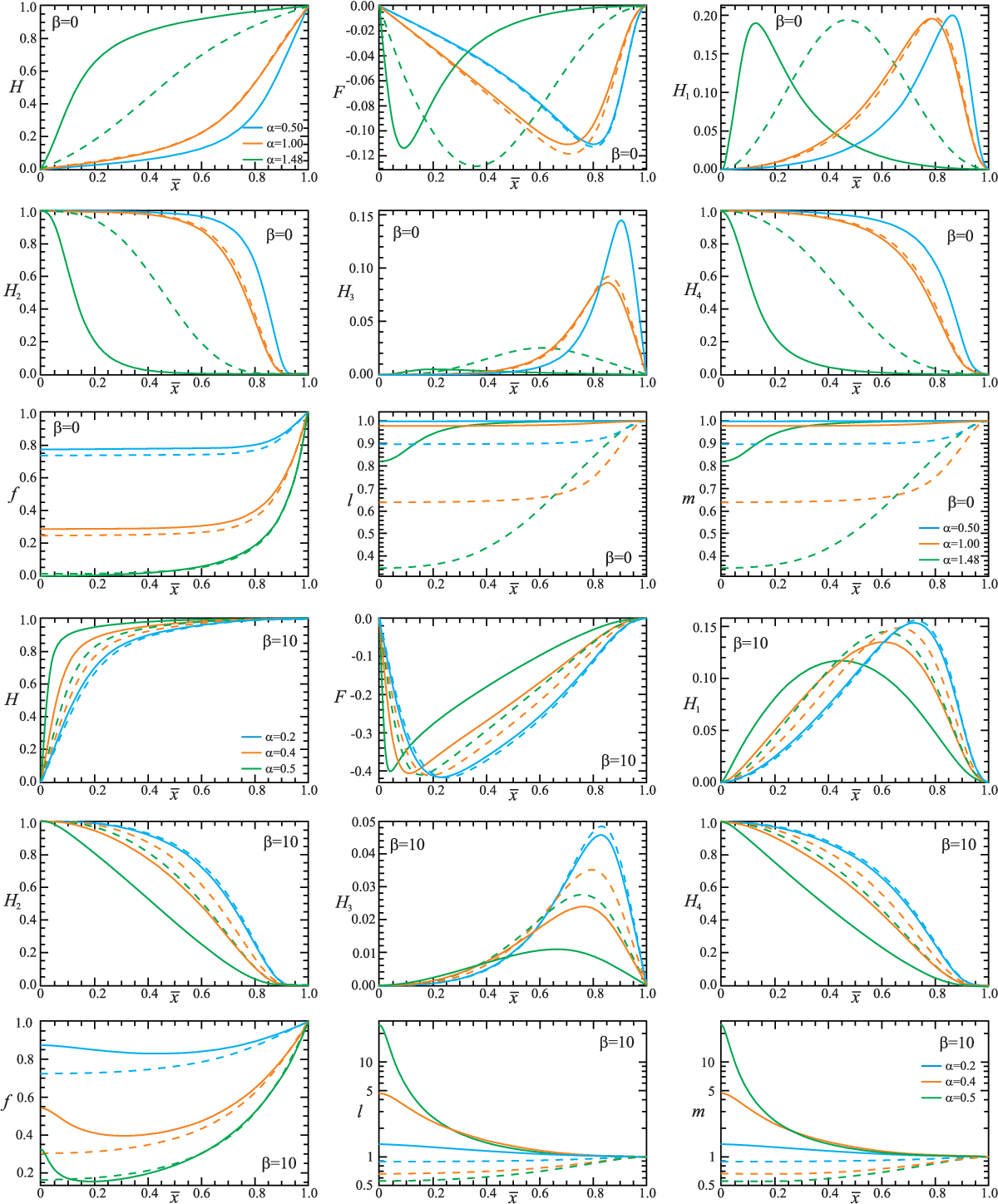}
\end{center}
\vspace{-0.5cm}
\caption{The axially symmetric  $n=2$ multimonopole solutions are shown  for $\beta=0, 10$. The solid lines correspond to GR ($\gamma=0$),
while the dashed lines represent MGT with $\gamma=-10$. The plots are made for a fixed angle $\theta=\pi/4$.
}
\label{fig_sols_n_2_beta_0_10}
\end{figure}


The numerical results for the spherically and axially symmetric cases are presented in Fig.~\ref{fig_mass_alph_n_1_2}.
To highlight the impact of the gravity modification, we compare the systems in GR ($\gamma=0$) and  MGT with
 $\gamma=-10$ for all possible values of  $\alpha$ and two  fixed values of  $\beta$ (0 and 10). 
(Note that we adopt a negative $\gamma$ because only in this case do the scalar curvature $R$ 
and the metric functions exhibit asymptotically damped behavior; for positive $\gamma$, 
these functions oscillate~\cite{Astashenok:2017dpo,Folomeev:2018ioy}.)
The graphs show that in the BPS limit, there is no noticeable difference in mass between the GR and MGT systems for either the
 spherically ($n=1$) and axially ($n=2$) symmetric cases. However, as  $\beta$ increases,
 the mass differences become significantly more pronounced, as shown in the right panel of Fig.~\ref{fig_mass_alph_n_1_2} for $\beta=10$.
 This effect is particularly evident for the axially symmetric configurations, where the mass difference at a fixed $\alpha$
 can reach $\sim 15 \%$. Furthermore, for fixed $\alpha$ and $\beta$, the mass of (multi)monopoles 
in  MGT is always smaller than that in GR.
(In this regard, a similar situation occurs for neutron stars in MGT as well~\cite{Folomeev:2018ioy}.)
Finally, just as in GR, both monopole and multimonopole solutions in  MGT converge to the corresponding extremal RN black hole solution in the limit
$\alpha\to \alpha_{\text{cr}}(n,\beta)$.

It is seen from Fig.~\ref{fig_mass_alph_n_1_2} that with increasing  $\alpha$, the mass {\it per unit charge} of 
the configurations under consideration always decreases. In the BPS limit, for $\alpha=0$, the mass of the $n=2$ multimonopole
{\it per unit charge} is precisely equal to the mass of the $n=1$ monopole. 
However, for $\alpha>0$, the mass {\it per unit charge} of the multimonopoles is smaller than that of the $n=1$ monopole. 
As pointed out in Ref.~\cite{Hartmann:2001ic}, this is due to the fact that in the BPS limit in GR, there is an attractive phase (caused by gravitation) 
between like monopoles (i.e., between two monopoles with magnetic charges of like sign). 
In turn, the modification of GR  leads to an even greater reduction in mass with an increasing in the gravitational coupling.
Moreover, the multimonopoles exist for gravitational coupling strengths larger than those possible for the $n=1$ monopoles. 

In turn, when  $\beta\neq 0$, the situation is more complicated~\cite{Hartmann:2001ic}. 
The flat-space and small-$\alpha$ multimonopoles have a higher mass {\it per unit charge} than the $n=1$ monopole
(reflecting a repulsive phase between like monopoles). However, for larger $\alpha$, this repulsive phase can give way to an attractive one;
i.e., the repulsion between like monopoles can be overcome, for a small Higgs self-coupling, by sufficiently strong gravitational attraction.
Then there is some equilibrium value $\alpha_{\text{eq}}$ for which the multimonopole mass {\it per unit charge} 
and the monopole mass are equal to one another.  An increase in $\beta$ leads to a growth of $\alpha_{\text{eq}}$, 
and for large values of the Higgs self-coupling, the equilibrium value $\alpha_{\text{eq}}$ is no longer present. 
This is precisely what is observed in the right panel of Fig.~\ref{fig_mass_alph_n_1_2}:
at $\beta=10$, the multimonopole mass {\it per unit charge} never equals the mass of the $n=1$ monopole.
However, it is possible to obtain multimonopoles in  MGT whose mass {\it per unit charge} becomes equal to the mass of the $n=1$ monopole in GR,
even at large $\beta$ (see the right panel of Fig.~\ref{fig_mass_alph_n_1_2} where the corresponding $\alpha_{\text{eq}}\approx 0.75$).
Furthermore,  for the Higgs self-coupling $\beta\gg 1$, 
the repulsive phase dominates; consequently, in GR, the multimonopole mass is always significantly greater than the mass of  the $n=1$ monopoles. 
 However, in contrast to GR, this effect is somewhat attenuated  in MGT. For large~$\alpha$, the mass of the $n=2$ multimonopoles 
approaches the mass of the $n=1$ monopoles, as shown in the right panel of Fig.~\ref{fig_mass_alph_n_1_2}.
Notably,  in contrast to the BPS limit, the multimonopoles at sufficiently large $\beta$ exist for gravitational coupling strengths smaller 
than those possible for the $n=1$ monopoles. 

Notice that the numerical calculations indicate that a further increase in $|\gamma|$ up to $\gamma=-1000$ has a negligible effect on the mass.
This is apparently related to the fact that the scalar curvature satisfies $|R|\ll 1$ for all systems under consideration, 
and an increase in $|\gamma|$ is accompanied by a corresponding decrease in $R$.
Consequently, the contribution from the nonlinear term in  MGT is compensated by the smallness of $R$.

The results of numerical calculations are illustrated in  Figs.~\ref{fig_sols_n_1} and \ref{fig_sols_n_2_beta_0_10}, 
which show typical spatial distributions of the field functions for various system parameters.
These graphs indicate that differences in the behavior of the field functions generally grow as the coupling constant $\alpha$ increases.
This, in turn, explains why the mass difference between the systems under consideration increases with $\alpha$ (see Fig.~\ref{fig_mass_alph_n_1_2}).
The inclusion of the Higgs self-coupling ($\beta\neq 0$) significantly deforms the solutions and leads to qualitative changes in their behavior. 
For example, at $\beta=0$, the metric function $f$ increases monotonically with $x$ for all $\alpha$. Its minimum value occurs  at the origin ($x=0$) 
and decreases as $\alpha$ approaches the critical value  $\alpha_{\text{cr}}$, where $f(0)\to 0$.
 However, at $\beta=10$ in GR, the minimum of $f$ is located at $x\neq 0$,  though as $\alpha\to \alpha_{\text{cr}}$, the minimum value again tends to zero at $x\to 0$.
  In MGT, the minimum remains at $x=0$, regardless of $\alpha$. 
 Furthermore, at $\beta=0$, the metric functions $l$ and $m$   are always less than unity in both GR and MGT.
At $\beta=10$, however, these functions significantly exceed 1 in GR, while remaining less than 1 in MGT.

Consider now the behavior of the gauge functions. For a fixed $\beta$, 
the gauge functions $W$ (spherically symmetric case) and $H_2, H_4$ (axially symmetric case) demonstrate
similar qualitative behavior  (with corresponding deformations at $\beta=10$). 
In turn,  the function $H_1$ possesses a maximum whose position shifts toward the origin as $\alpha$ increases;
this shift occurs noticeably more slowly in MGT than in GR. Furthermore, while the height of the maximum depends only weakly 
on $\alpha$ at $\beta=0$, a significant $\alpha$-dependence is observed at $\beta=10$, especially in GR.
The function $H_3$ also has a maximum, the position and height of which decrease with increasing $\alpha$,
with MGT showing a slower rate of decrease than GR.
At the same time, the calculations indicate that the region where the gauge fields differ considerably from zero shrinks as $\alpha$ 
increases and vanishes as  $\alpha\to\alpha_{\text{cr}}$.

For $\beta=0$, the Higgs field $H$ increases monotonically with $x$. The field $F$
possesses a minimum with a negative value, the position of which shifts toward the origin (decreases) as $\alpha$ increases;
this shift occurs more slowly in MGT than in GR. 
Again, the height of the extremum depends only weakly on  $\alpha$. 
At $\beta=10$, the behavior of the solutions is generally similar, 
though they become more concentrated toward the center due to the scalar field self-interaction:
a nonzero Higgs mass $M_s$ tends to force the scalar field rapidly toward its vacuum value.
This alters the monopole core structure and increases field gradients and magnetic concentration in the core region, 
further raising the energy and increasing the ADM mass of both monopoles and multimonopoles (see the right panel of Fig.~\ref{fig_mass_alph_n_1_2}).
In turn, the modification of gravity results in a less compressed monopole core,
as is clearly seen from the behavior of the Higgs field $H$ at large values of the gravitational coupling constant $\alpha$.

\section{Conclusions}

Non-Abelian monopoles are objects whose structure and physical characteristics are largely determined 
not only by the parameters of the Yang-Mills and scalar fields, but also by their own strong gravitational fields. 
Within GR,  such monopoles have been extensively studied in various formulations, 
and it has been demonstrated that there exist entire families of new solutions, including black hole solutions with a regular event horizon.
However, monopole solutions in modified gravity have received considerably less attention.

This paper aims to fill this gap by considering static spherically and axially symmetric (multi)monopole solutions 
within modified gravity (specifically, the Starobinsky model). For this purpose, 
we have numerically constructed solutions of the full sets of coupled field equations  and investigated their properties. 
To reveal the effects of gravity modification on the physically relevant characteristics of monopoles, we have investigated the dependence of their mass on
the effective gravitational coupling constant $\alpha$ at fixed values of  the Higgs self-coupling constant $\beta$ 
and the modified gravity parameter $\gamma$. 

Let us outline the most interesting features of the configurations under consideration: 
\begin{itemize}
\item[(i)] It is shown that for all configurations considered, the modification of gravity leads to a reduction in mass.  This effect becomes particularly pronounced at large values of
  $\beta$ and for the multimonopole solutions, where the mass difference between monopoles in GR and  MGT can reach $\sim 15\%$.

\item[(ii)] It is found that the modification of gravity enables the existence of solutions at substantially larger values of
the gravitational coupling constant $\alpha$. In this case, as in GR, there is a critical value of $\alpha\to\alpha_{\text{cr}}$
for which both monopole and multimonopole solutions  converge to the corresponding extremal RN solution.

\item[(iii)] It is demonstrated that in  MGT, as in GR, for $\beta\gg 1$, the mass of the $n=2$ multimonopoles is noticeably larger than that of the 
$n=1$ monopoles. However, in  MGT this effect is somewhat attenuated;  for large $\alpha$, the mass of the $n=2$ multimonopoles approaches the mass of the $n=1$ monopoles.

\item[(iv)]  It is found that a further increase in the modified gravity parameter $|\gamma|$ has a negligible effect on the $(M-\alpha)$ dependence.
\end{itemize}
 
A small change in the $(M-\alpha)$ dependence, together with a substantial increase in $|\gamma|$, suggests that even for very large values of 
the parameter $|\gamma|$~-- used, for example, in modeling neutron stars~\cite{Astashenok:2017dpo,Folomeev:2018ioy}~-- the monopole characteristics 
will remain practically the same as in the case of $\gamma=-10$ considered in the present work. It may also be expected that, within other $f(R)$ theories of gravity 
(for instance, those with exponential, logarithmic, or cubic corrections in $R$ employed in modeling compact stellar objects~\cite{Olmo:2019flu}), 
the situation will be qualitatively and quantitatively similar to that obtained here for the model with a quadratic correction in $R$. 
Nevertheless, this issue requires special investigation, as does the question of whether other types of modified gravity 
could  yield qualitatively new solutions describing gravitating monopoles with significantly different 
physical properties compared to those in GR.

\end{document}